\documentclass[twocolumn,showpacs,preprintnumbers,amsmath,amssymb]{revtex4}

\usepackage{graphicx}
\usepackage{dcolumn}
\usepackage{bm}

\begin{document}

\title{Delayed feedback induced directed inertia particle transport in a washboard potential}

\author{D. Hennig}
\author{L. Schimansky-Geier}
\affiliation{Institut f\"{u}r Physik, Humboldt-Universit\"{a}t
 zu Berlin, Newtonstr.~15, 12489 Berlin, Germany}
\author{P. H\"anggi}
\affiliation{Institut f\"{u}r Physik, Universit\"{a}t
  Augsburg, Universit\"atsstr.~1, 86135 Augsburg, Germany}

\date{\today}

\begin{abstract}
\noindent We consider motion of an underdamped Brownian particle
in a washboard potential that is subjected to an unbiased time-periodic external
field.
While in the limiting deterministic system in dependence of the strength and phase of the external field directed net motion can exist, for a finite temperature the net motion averages to zero.
Strikingly, with the application of an additional time-delayed
feedback term directed particle motion can be accomplished persisting up to
fairly high levels of the thermal noise.  In detail, there exist values of the feedback strength and delay time for which the feedback term performs oscillations that are phase locked to the time-periodic external field. This yields an effective biasing rocking force promoting
periods of forward and backward motion of distinct duration, and thus directed motion. In terms of phase space dynamics
we demonstrate that with applied feedback desymmetrization of
coexisting attractors takes place leaving the ones supporting either positive or negative velocities as the only surviving ones. Moreover, we found parameter ranges for which in the presence of thermal noise the directed transport is enhanced compared to the noise-less case.

\end{abstract}

\pacs{05.40.-a, 02.50.Ey, 05.60.-k, 05.65.+b, 87.15.-v}
\maketitle



\section{Introduction}
Transport phenomena play a fundamental role in many physical
systems. In this context, the theme of a ratchet dynamics has
attracted considerable interest over the past years. This is
particularly due to the fact that the ratchet effect assists the
creation of a directed flow of particles without the presence of any
bias force in the system. The ratchet dynamics has been mainly
applied to biological or mesoscopic systems where Brownian motion in
a periodic {\it asymmetric} potential together with dissipation
enters in some form the problem and the directed motion is generated
from nonequilibrium noise, see the various overviews on molecular
and Brownian motors in Refs. \cite{Ha1996}-\cite{RMP}. On the other
hand for periodic systems with maintained spatial symmetry the
accomplishment of directed net motion necessitates that the system
is exerted to additional biasing (symmetry-reducing) impacts. Here
we consider particle motion in a washboard potential which is often
employed as as a paradigm to model transport in one-dimensional
periodic and {\it symmetric} structures
\cite{Risken},\cite{Reguera}-\cite{Hennig}. A symmetric (unbiased)
external force is assumed to rock the potential. Our aim is to
demonstrate that directed particle transport can be achieved with
the application of a time-delayed feedback method in a wide
temperature range. Although the delayed feedback method was
originally proposed by Pyragas \cite{Pyragas} to stabilize unstable
states in deterministic systems meanwhile it has been facilitated in
various other contexts \cite{handbook} among them there is also the
control of purely noise-induced oscillations \cite{Janson},\cite{Prager}.
Recently
in the context of controlling transport in Brownian motors a
feedback strategy has been successfully utilized for two ratchet
systems interacting through a unidirectional delay coupling
\cite{delay}. The effect of time-delayed feedback on the
rectification of thermal motion of  Brownian particles has been
studied in overdamped ratchet systems \cite{Cao}-\cite{Wu}. A recent
experimental implementation  using such a feedback mechanism for a flashing ratchet has been realized with an optical line trap: it has been observed
that the use of feedback increases the ratchet velocity up to an
order of magnitude \cite{Lopez}, in agreement with theory.

Stabilization of chaotic motion in deterministic inertia ratchet
systems to increase the current efficiency was considered in Refs.
\cite{Family},\cite{Son}. Furthermore, an asymmetric ratchet
potential with included time-delayed feedback was treated in the
context of an inertial Brownian motor \cite{Wu1}.

Our paper, dealing with time-delayed feedback induced directed
motion, is organized as follows: First we introduce the model of an
inertial Brownian particle evolving in a symmetric spatially
periodic potential under the influence of an additional time-delayed
feedback term. The subsequent section concerns the underlying
deterministic dynamics. In particular bifurcation diagrams with and
without applied time-delayed feedback are discussed. The impact of a
heat bath of fixed temperature on the particle transport features is
studied in Section \ref{section:thermal}. Finally we summarize our
results.

\section{The model}
We consider an inertia Brownian particle that is moving along a
one-dimensional periodic structure. The dynamics is governed by
the following inertial Langevin equation expressed in
dimensionless form
\begin{eqnarray}
\ddot{q}+\gamma \dot{q}&=&-\frac{dU}{dq}+F\sin({\omega
\,t+\theta_0})+\xi(t)+f(t)\,.\label{eq:qdot}
\end{eqnarray}
The dot denotes differentiation with respect to time. The particle
evolves in a spatially-periodic and {\it symmetric} potential
\begin{equation}
U(q)=U(q+1)=-\cos(2\pi q)/(2\pi)\,,
\end{equation}
 of unit period  $L=1$ and
barrier height $\Delta E=1/\pi$ and its position and velocity are
quantified by the variable $q(t)$ and $\dot{q}(t)\equiv v(t)$
respectively. The particle is driven by an external,
time-dependent forcing field of amplitude $F$, frequency $\omega$
and phase $\theta_0$. In addition it is subjected to a Gaussian
distributed thermal, white noise $\xi(t)$ of vanishing mean
$\langle\xi(t)\rangle=0$, obeying the well-known
fluctuation-dissipation relation $\langle\xi(t)
\xi(t^{\prime})\rangle=2\gamma k_B T\delta(t-t^{\prime})$ with
$k_B$ and $T$ denoting the Boltzmann constant and temperature, 
respectively. The friction strength is measured by the parameter
$\gamma$. The last term in Eq.~(\ref{eq:qdot}) denotes a continuous
time-delayed feedback term of the form
\begin{equation}
f(t)=K[\dot{q}(t-\tau)-\dot{q}(t)]\label{eq:feedback}
\end{equation}
of strength $K$ and with delay time $\tau$. Before we embark on
the study of the Brownian particle motion it illustrative to
consider the deterministic limiting case arising for vanishing
thermal noise, i.e. $T=0$. For our study we fix the following parameter
values $\gamma=0.1$, $\omega=2.25$, $\theta_0=0$.

\section{The deterministic case}
\begin{figure}
\includegraphics[scale=0.35]{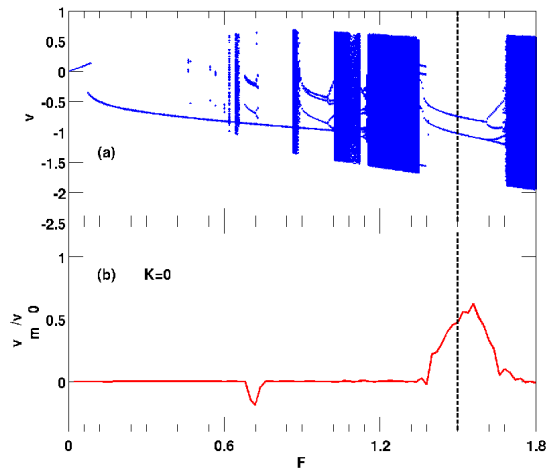}
\caption{\label{fig:bif1} (a): Bifurcation diagram as a function
of the amplitude of the external driving without time-delayed
feedback, i.e. $K=0$ and remaining parameter values:
$\omega=2.25$, $\gamma=0.1$ and $T=0$.
(b): Mean velocity $v_m/v_0$ as a function of the amplitude of the external
driving. For later use a dashed vertical line is drawn at the value $F=1.5$}
\end{figure}

The dynamics of the deterministic system ($T=0$) exhibits very
rich and complex behavior and depending on the parameter values
and initial conditions one finds periodic, or aperiodic
(quasiperiodic and/or chaotic) solutions in the long time limit
\cite{Baker}-\cite{Kostur}. The character of the phase flow
evolving without feedback term ($K=0$) in a three-dimensional phase space is
conveniently displayed by a Poincar\'{e} map using the period of
the external force, $T_e=2\pi/\omega=2\pi/2.25\simeq 2.791$, as the
stroboscopic time. The deterministic equation of motion was
integrated numerically and omitting a transient phase points were
set in the map at times being multiples of the period duration
$T_e$. In Fig.\,\ref{fig:bif1}\,(a) the bifurcation diagram as a
function of the amplitude of the external driving is depicted.

Particle transport is quantitatively assessed by the mean velocity which we
define as the time average of the ensemble averaged velocity, i.e.
\begin{equation}
v_m= \frac{1}{T_s}\,\int_0^{T_s} dt^{\prime} \langle{v}_n(t^{\prime})\rangle \,,
\end{equation}
with simulation time $T_s$ and with the ensemble average given by
\begin{equation}
\langle {v}_n(t)\rangle=\frac{1}{N}\sum_{n=1}^N\,{v}_n(t)\,.\label{eq:ensemble}
\end{equation}
Here $N$ denotes the number of particles constituting the ensemble with associated initial conditions $q_n(0)$ and $v_n(0)$ that are chosen such that
the interior of the unperturbed separatrix in phase plane is uniformly covered.
We express $v_m$ in terms of the ratio of the spatial and temporal
periods $L/T_e \equiv v_0$ with $v_0\simeq 0.358$ being the
velocity for running solutions that advance by one spatial period
during one period duration of the external field.
\begin{figure}
\includegraphics[scale=0.35]{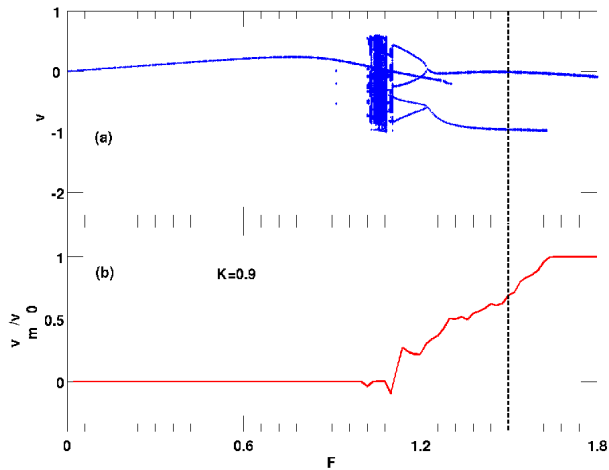}
\caption{\label{fig:bif2}  As in Fig.~\ref{fig:bif1} but with
switched-on  time-delayed feedback of strength $K=0.9$ and delay
time $\tau=1.95$. The remaining parameter values read as: $\omega=2.25$,  $\gamma=0.1$ and $T=0$.}
\end{figure}

For undercritical amplitudes of the modulation force $F\lesssim
0.6$ those particles which are initially residing near the bottom of a
potential well, remain trapped. Increasing $F$ leads to escape
from the potential wells and the particle jumps subsequently from
one well to another one. The arising two typical scenarios are the
pinned and running states respectively. In the former state the
motion proceeds at most over a finite number of spatial periods
whereas in the latter state motion is directed and unrestricted in
the spatial dimension. In terms of the phase flow running
asymptotic solutions correspond to phase locked periodic attractors transporting a particle with velocity
$v=m/n$ over $m$ spatial periods of the potential during $n$
period durations $T_e$ of the external periodic field. Running asymptotic solutions may also be supported by aperiodic attractors.

In the bifurcation diagram associated with the dynamics without
applied time-delayed feedback shown in  Fig.\,\ref{fig:bif1}\,(a)
one recognizes vertically extended stripes covered densely with
points corresponding to non phase locked aperiodic attractors and
several periodic windows as well as period-doubling cascades to
chaos. These features of the phase flow are readily attributed to
the resulting mean velocity of the net motion (depicted in
Fig.~\ref{fig:bif1}\,(b)).
The ensemble average is taken over an ensemble of $N=5000$
trajectories with uniformly distributed initial conditions $q(0)$ and $v(0)$.
For computation of the long-time average the simulation time
interval for each trajectory is taken as $T_s=5\times 10^5 \simeq 1.8\times 10^5\times
T_e$. We notice almost in the entire $F$-range vanishing mean
velocity $v_m=0$. The exceptions are the intervals $0.74\lesssim F
\lesssim 0.80$ and $1.36\lesssim F \lesssim 1.62$ for which the
solutions are associated with multiple coexisting attractors lying
in fairly extended periodic windows in Fig.~\ref{fig:bif1}(a).
Focusing interest on the latter one we note that at $F\gtrsim
1.36$ tangent bifurcations give birth to two coexisting period-one
attractors. The upper one of them is related with positive
particle velocity $v=v_0>0$ whereas on the lower one particles
move with velocity $v=-v_0<0$. For increasing $F\gtrsim 1.62$ these
attractors are destroyed by way of crisis after passage through a
period-doubling route to chaos. 

The oppositely running solutions
attributed to the two period-one attractors contribute to the mean
velocity with different weight with the one with positive velocity $v=v_0$ being
{\it dominant} and thus yielding the window of positive mean
velocity $v_m$. Apparently, the directed motion results from a
lowering of the dynamical symmetry caused by the external
modulation field \cite{Yevtushenko},\cite{Hennig}. That is, even though the
potential and the external modulation field are spatially and
temporarily symmetric respectively, with the choice of a  fixed
phase $\theta_0$ the symmetry of the flow is reduced and a
phase-dependent net motion is found. (Note that additional
averaging over the phase $\theta_0$ yields zero mean velocity.)
Due to symmetry reasons it holds that the sign of the mean
velocity is reversed upon the changes $\theta_0=0\,\rightarrow
\theta_0=\pi$ and $F\rightarrow -F$ respectively. However, there
exists a phase $0<\theta_0<\pi$ for which symmetry between the two
coexisting periodic attractors supporting solutions with
velocities of opposite sign, $v_0$ and $-v_0$, is restored and
therefore the net motion vanishes.

In the numerical simulation of the system (\ref{eq:qdot}) with applied time-delayed feedback term (\ref{eq:feedback}) we set $f(t)=0$ in the interval
$t\in [0,\tau)$, that is the system is affected by $f(t)$ only for $t\geqq\tau$.
We performed extensive numerical studies to identify optimal
parameters of the feedback term which establish efficient
directed net motion. It turns out that this is achievable for
delay times in the range of $1.65 \lesssim \tau \lesssim 2.00$ and
for feedback strength $K\gtrsim 0.8$ (see also further below in Fig.~\ref{fig:current}). In the following
we illustrate exemplarily the impact of time-delayed feedback on
the transport properties for a feedback strength $K=0.9$ and delay
time $\tau=1.95 \simeq 0.7 \times T_e$. With such appropriate feedback term
applied the extension of the aperiodic regions shrinks
considerably and only a comparatively narrow band of aperiodic
behavior for $1\lesssim F \lesssim 1.11$ prevails in the
bifurcation diagram illustrated in Fig.~\ref{fig:bif2}\,(a).
Remarkably, for $F\gtrsim 1.63$ the lower period-one attractor
supporting negative velocities looses stability and is converted
into a repellor leaving the upper attractor of positive velocity
as the only persisting attractor. Thus application of feedback
results in a reshaping of the bifurcation diagram. In fact,
due to the absence of the period-one attractor supporting motion with velocity $v=-v_0$ only running
solutions with velocity $v_m=v_0$ are then recognized in
Fig.~\ref{fig:bif2}\,(b) showing the mean velocity as a function
of $F$. Otherwise the mean velocity raises form zero level for
overcritical $F\gtrsim 1.1$ and grows with increasing $F$ until
$F\gtrsim 1.63$ when $v=v_0$ is attained.

\begin{figure}
\includegraphics[scale=0.3]{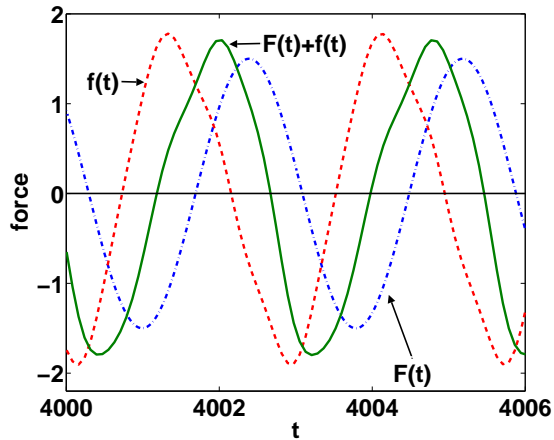}
\caption{\label{fig:control}  Time evolution of the feedback term
$f(t)$, the driving force term $F(t)$ and their sum $F(t)+f(t)$ with $K=0.9$ and delay time $\tau=1.95$. The remaining parameter values read as: $F=1.8$, $\omega=2.25$,  $\gamma=0.1$ and $T=0$.}
\end{figure}
In order to gain insight into the feedback-induced mechanism that leads to directed transport (occurring for $F\gtrsim 1.63$ in the period-one window in Fig.~\ref{fig:bif2}) we display the temporal behavior of the
feedback term $f(t)$, given in Eq.~(\ref{eq:feedback}), and the external driving term $F(t)=F\sin(\omega t)$ and their sum $F(t)+f(t)$ in Fig.~\ref{fig:control} for driving amplitude $F=1.8$.
Throughout the time the feedback term $f(t)$
 performs oscillations possessing  considerable asymmetry.
Most importantly, these oscillations are entrained to the (symmetric) external driving term with a phase shift.
The sum $F(t)+f(t)$, performing asymmetric oscillations, determines the {\it effective rocking force} exerted on the particle. This ratching force is self-induced due to the feedback in comparison with the externally imposed ratching force in the form of asymmetric periodic driving fields considered in  \cite{Yevtushenko}.

\begin{figure}
\includegraphics[scale=0.3]{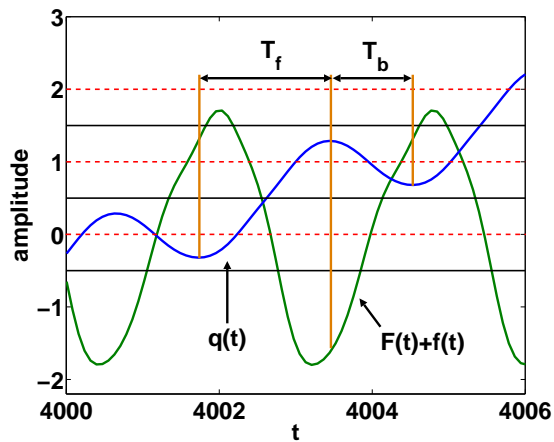}
\caption{\label{fig:amplitude}  Time evolution of the particle amplitude $q(t)$ and the effective rocking force $F(t)+f(t)$ with $K=0.9$ and delay time $\tau=1.95$. The remaining parameter values read as: $F=1.8$, $\omega=2.25$, $\gamma=0.1$ and $T=0$. Horizontal solid (dashed) lines indicate the position of the maxima (minima) of the unbiased washboard potential.}
\end{figure}

How the directed rightward particle motion is enforced by this effective biasing
rocking force is illustrated in Fig.~\ref{fig:amplitude}. (For the present discussion we discard the contribution from $-dU(q)/dq$ to the total force.)
Clearly, an oscillating rocking force leads to passages of forward and backward motion, also called enhancement and depreciation periods. Crucially,
the feedback term is suitably entrained to the external driving term in such a way that
the period of forward motion is longer than its backward counterpart, denoted by $T_f$ and $T_b$ respectively in Fig.~\ref{fig:amplitude}. To be precise, at the moment when the backward motion of the particle terminates  the effective rocking force has not yet reached its maximum as indicated by the left vertical line in Fig.~\ref{fig:amplitude}.
Subsequently the particle moves in the forward direction due to the ongoing positive rocking force that after passing through its maximal value declines. Nevertheless, until half of the time span $T_f$ is reached the particle motion is  still enhanced in the forward direction. Afterwards for $t>T_f/2$ the effective rocking force becomes negative and
thus the momentum of the particle is reduced steadily with increasingly negative values of $F(t)+f(t)$.  The end of
the forward-motion period is designated by the middle vertical line in Fig.~\ref{fig:amplitude}.
However, there remains only
comparatively little time, namely $T_{b}<T_f$,
during which backward-motion is enforced, that is when during the depreciation period the effective rocking force is negative.

Consequently, the
asymmetry in the enhancement and depreciation phases serves for a rather long period of forward-motion compared to the backward-motion. Therefore the
effective motion of the particle proceeds to the right. Notably, this feature is induced by entrainment of the asymmetric time-delayed feedback term to the symmetric external modulation field if the feedback strength and delay time are suitably chosen.
Notice that this oscillation behavior of the time-delayed feedback term is
different from noninvasive control methods where the delayed feedback control
vanishes once a targeted unstable periodic orbit has been stabilized \cite{Pyragas},\cite{handbook},\cite{Son},\cite{Wu1}.

\section{Directed thermal net particle motion}\label{section:thermal}
We now study the impact of a heat bath of fixed temperature $T>0$ on the
particle transport features. With the inclusion of finite thermal
noise transitions between the now metastable attractors are likely
and, independent of the initial conditions, trajectories permeate
the whole phase space rendering the dynamics ergodic.

\begin{figure}
\includegraphics[scale=0.4]{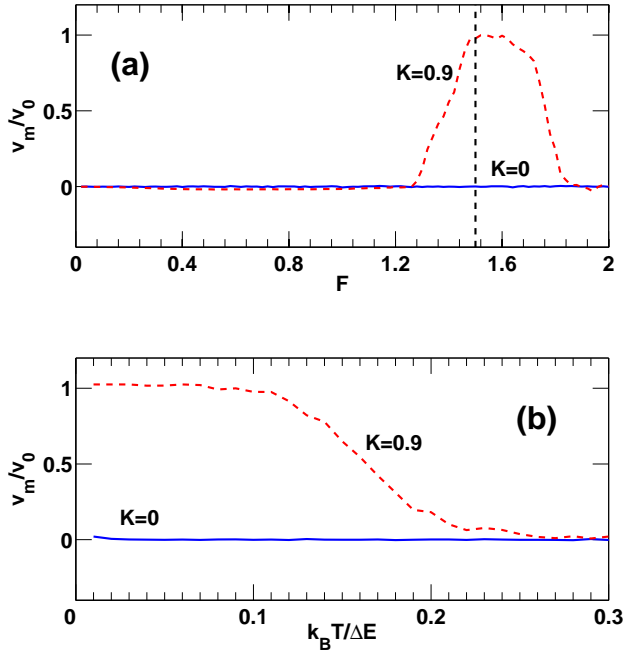}
\caption{\label{fig:currentA} (a): Mean velocity in
dependence of the external modulation field amplitude without and
with applied time-delayed feedback force as indicated in the plot and for 
thermal energy $k_BT=0.1\times \Delta E$.
(b): Mean
velocity in dependence of the thermal energy expressed in units of
the barrier energy, $k_BT/\Delta E$, and for fixed $F=1.5$ and delay time $\tau=1.95$.
The remaining parameter values are $\omega=2.25$ and $\gamma=0.1$.}
\end{figure}

For the computation of the mean velocity the ensemble average in (\ref{eq:ensemble}) was taken over $N=5000$ realizations of the
thermal noise for an arbitrarily chosen initial condition.
The mean velocity $v_m$
represented without feedback term but in the presence of thermal noise of a fairly high
level of $k_BT=0.1 \times \Delta E$ in Fig.~\ref{fig:currentA}\,(a)
is zero regardless of the value of the modulation field
strength $F$. This has to be distinguished from the preceding
noise-less case where there exist even for $K=0$ regions of
nonvanishing net motion (cf. Fig.~\ref{fig:bif1}) due to the fact
that motion on attractors with different sign of the velocity contribute
with distinct weight to the asymptotic net motion. In other words,
the impact of the noise leads to symmetrization of the basins of
attraction of transporting periodic and/or aperiodic attractors. Therefore any initial condition yields zero asymptotic current.

Interestingly, this situation changes imposing the Langevin
dynamics additionally to the time-delayed feedback and we found parameter
constellations for which directed net motion results despite the
presence of strong noise.

As Fig.~\ref{fig:currentA}\,(a) reveals,
applying feedback of strength $K=0.9$ and delay time $\tau=1.95$,
the mean velocity as a function of the external modulation field
strengths exhibits a resonance-like structure for $1.2\lesssim F
\lesssim 1.7$,  i.e. for values of $F$ for which transport exists
in the deterministic case. Strikingly, with the impact of thermal noise the
feedback-controlled transport proceeds more efficient in
comparison with the deterministic case in the fairly wide range $1.2 \lesssim F\lesssim 1.7$
(cf. Figs.~\ref{fig:bif2} and \ref{fig:currentA}\,(a)). On the other hand, for $F\gtrsim 1.7$ the directed transport feature of the deterministic system is destroyed by the thermal fluctuations.
\begin{figure}
\includegraphics[scale=0.3]{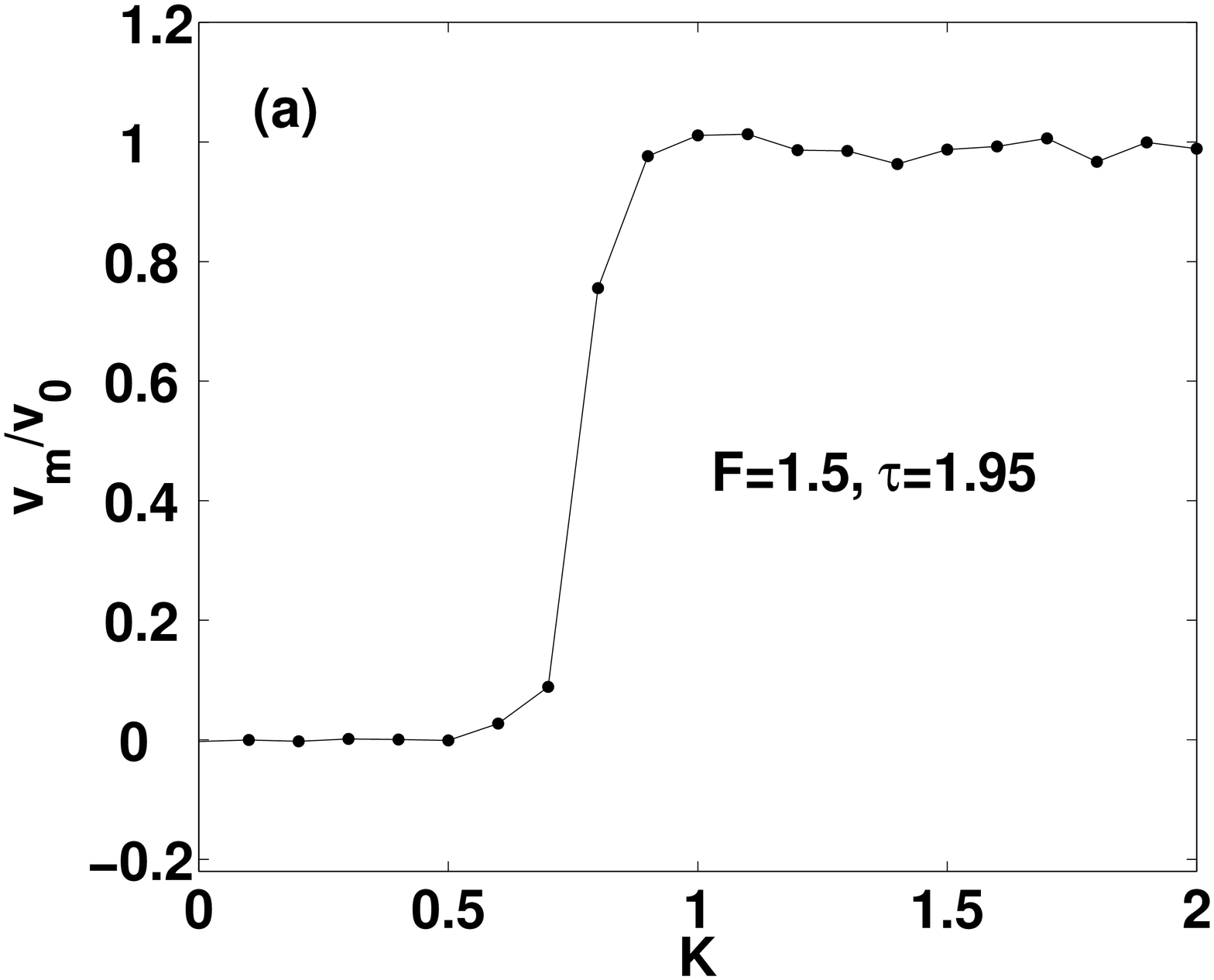}
\includegraphics[scale=0.3]{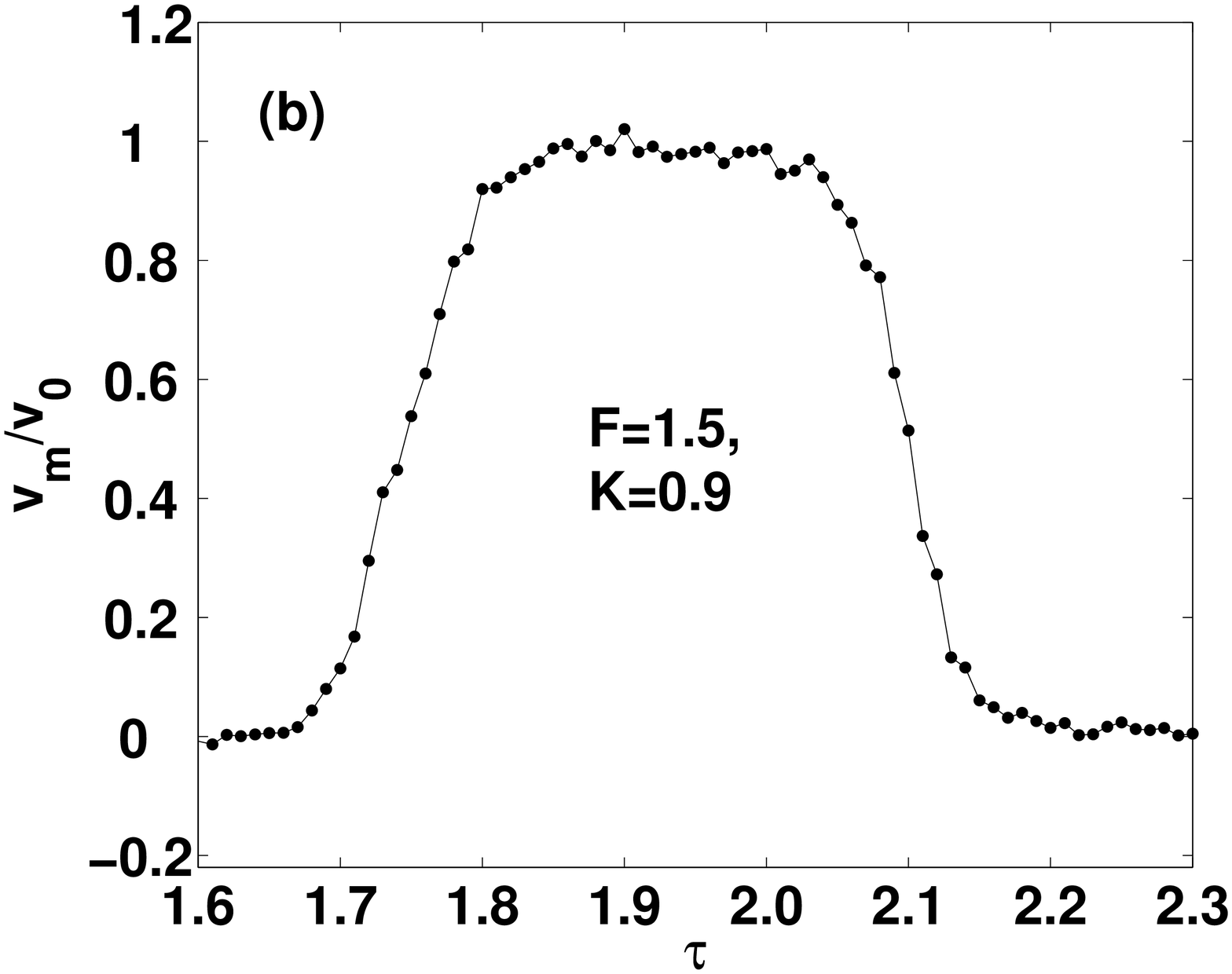}
\caption{\label{fig:current}
(a) Mean velocity in
dependence of the strength of the time-delayed feedback and for fixed
 $F=1.5$ and $\tau=1.95$. (b): Mean velocity in dependence
of the delay time and for fixed $F=1.5$ and $K=0.9$. The
remaining parameter values are $\omega=2.25$, $\gamma=0.1$ and
$k_BT=0.1 \times \Delta E$.}
\end{figure}
Furthermore, like in the deterministic case, there exist a threshold value for the feedback
strength beyond which directed net motion is achieved (see  Fig.~\ref{fig:current}\,(a)) and the range of delay times
being optimal for running solutions is indicated by the
resonance-like structure in
Fig.~\ref{fig:current}\,(b).

>From Fig.\,\ref{fig:currentA}\,(b),
illustrating the mean velocity in dependence of the noise
strength, one infers that directed transport is sustained up to
comparatively high noise strength before it ceases eventually to
exist. On the other hand, it is seen that without feedback term, i.e. $K=0$,
the mean velocity averages to zero already for low noise intensity.
As the damping strength is concerned we found that feed-back-induced directed motion is maintained as long as the system remains in the underdamped regime, that is for $\gamma \lesssim 2.5\simeq \omega_0$ with $\omega_0=\sqrt{2\pi}$ being the frequency of harmonic oscillations near the bottom of a potential well.  

\begin{figure}
\includegraphics[scale=0.35]{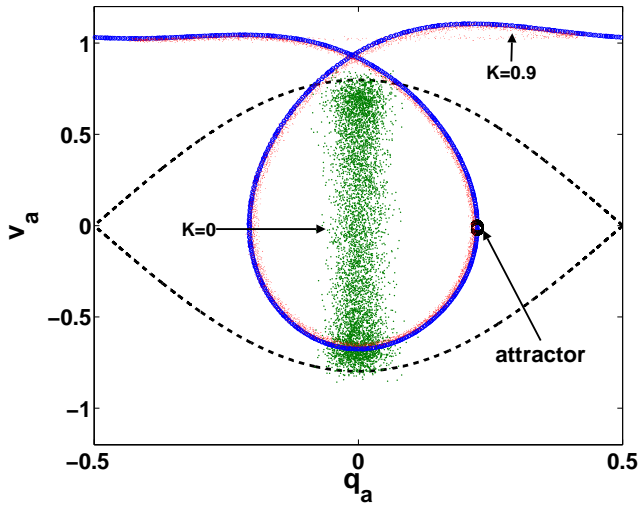}
\caption{\label{fig:sepesc} Stroboscopic map of the average
trajectory without ($K=0$) and with applied feedback ($K=0.9$ and
$\tau=1.95$) as indicated in the plot. The remaining parameter
values read as $F=1.5$, $\tau=1.95$ and $k_BT=0.1\times \Delta E$. The attractor with
$v=v_0>0$ of the deterministic system arising with the applied
feedback is also shown. The dashed lines indicate the separatrix
of the deterministic conservative undriven system, i.e.
$\gamma=F=0$.
We also superimposed the projection of the orbit of the
deterministic system on the $v-q-$plane (blue line).}
\end{figure}

\begin{figure}
\includegraphics[scale=0.3]{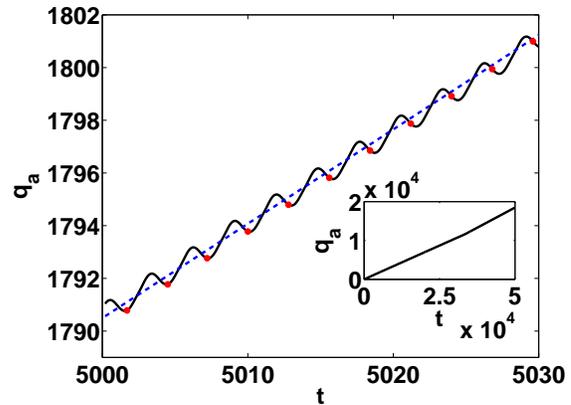}
\caption{\label{fig:Fig6} Time evolution of the average coordinate $q_a$ (wiggled line). The dashed straight possesses slope $v_0=L/T_e\simeq 0.358$. The bullets represent the position of the particle at moments being multiples of the period duration of the external modulation $T_e=2\pi/\omega$. The inset displays the long-time evolution of $q_a$. The parameter values read as $F=1.5$, $K=0.9$. $\tau=1.95$ and $k_BT=0.1\times \Delta E$.}
\end{figure}

Further insight into the origin of running solutions in the noise case is
gained from Fig.~\ref{fig:sepesc} showing the stroboscopic map of
the average trajectory defined as ${q_a}=\sum_{n=1}^N q_n/N$ and
${v_a}=\sum_{n=1}^N v_n/N$ with and without presence of the
time-delayed feedback for external modulation strength $F=1.5$, a
value for which in the deterministic case the mean velocity
remains nearly unaffected when the feedback is applied
(to ease the eyes the vertical
dashed line is drawn at the position of the value $F=1.5$
in Figs.\,\ref{fig:bif1},\ref{fig:bif2} and
\ref{fig:currentA}\,(a)).
Without feedback, $K=0$, the velocity range of the average
trajectory is not only symmetric with respect to $v_a=0$ but
remains confined within the boundaries of the separatrix loop of
the undriven, undamped deterministic system ($\gamma=F=0$). Thus
the solutions represent pinned states. With applied feedback of
strength $K=0.9$ the $v$-symmetry is broken. Moreover, the
stroboscopic plot of the average trajectory densely covers the
curve obtained when the periodic oscillations of the noise-less
system are projected onto the $q-v-$plane. This implies the existence
of an {\it attractive curve} related to near-torus-motion in phase space for the noise case.
The corresponding stroboscopic map of the dominant deterministic
periodic attractor with velocity $v=v_0$ is also drawn in Fig.~\ref{fig:sepesc}.
Crucially for directed net motion with $v_m=v_0$ arising for $T>0$ the average trajectory
sticks to the near-torus motion  never exploring other parts of
the phase space during the whole simulation time interval
$T_s=10^5$. The time evolution of the corresponding average coordinate $q_a$ is depicted in Fig.~\ref{fig:Fig6}.
Conclusively, despite the fact that the resulting
motion in the presence of noise and feedback evolves no longer perfectly
synchronous but still phase locked with the external periodic modulation it nevertheless
exhibits behavior reminiscent of that found in the corresponding
limiting deterministic system. To be precise the periodic oscillations of
the deterministic dynamics are replaced by near-torus motion in
the Langevin dynamics accomplishing phase locked aperiodic
transport in the sense that on average during one period duration
of the external field particles move by one spatial period, i.e. $v_m=v_0=L/T_e$.

\section{Summary}
In conclusion, we have identified and characterized a transport
regime for underdamped Brownian particles evolving in a {\it symmetric}
washboard potential under the mutual impact of an unbiased external periodic field and time-delayed feedback.
We have studied first the deterministic case of zero temperature.
It is has been shown that without feedback in some ranges of the amplitude and certain phases of the external modulation field windows of directed particle current exist.
This is due to the fact that attractors associated with phase locked oppositely-running solutions contribute with different weight to the net current. The direction of the net current can be reversed with a suitable choice of the phase of the external modulation field.

Interestingly with an applied time-delayed feedback in the deterministic system
there exist parameter ranges for which the attractors
supporting velocities of a definite sign loose stability and are converted
into  repellors leaving the  attractors of opposite velocity
as the only persisting ones. Hence, the efficiency of the net particle current is improved. There exist values of the feedback strength and delay time for which the feedback term performs oscillations that are phase locked to the time-periodic external field. This yields an effective biasing rocking force, promoting
periods of forward and backward motion of distinct duration, and thus directed motion.

On the other hand, for a finite temperature and without time-delayed feedback the net motion averages to zero. That is, the symmetry between the coexisting attractors supporting negative and positive velocities
is restored by the impacting thermal fluctuations.
Strikingly, in contrast to the case without feedback, with applied time-delayed feedback of appropriately chosen strength and delay time we find in a wide temperature range
complete desymmetrization of coexisting attractors supporting oppositely running solutions. As a consequence the particle motion proceeds  exclusively in one direction. Therefore the feedback-induced transport is not only robust with respect to thermal noise but moreover, there exist  parameter ranges for which in the presence of thermal noise the transport is more efficient than in the corresponding deterministic limiting
case. We have identified attracting curves in phase space which are linked with
motion on a torus with small deviations supporting transport.

We stress the difference between the transport control mechanism described above and the control of transport properties in inertia ratchet systems \cite{Family},\cite{Son},\cite{Wu1} facilitating the stabilization of certain targeted (unstable) periodic orbits via time-delayed feedback as well as the features in feedback flashing ratchets when the potential is alternatively switched on and off in dependence of the state of the system \cite{Cao}-\cite{Feito1}. In our case control of transport is achieved
if the time-delayed feedback term with appropriately chosen strength and delay time entrains to the external time-periodic field yielding effectively a biasing force that rocks the washboard potential such that distinct durations of the periods of forward and backward motion ensue.


\end{document}